\documentclass[
a4paper,%
12pt%
]{article}

\usepackage[dvips]{graphicx}

\title{New exactly solvable reflectionless potentials of Gamov's type}

\author{Sergei~P.~Maydanyuk
\thanks{E-mail: maidan@kinr.kiev.ua} \\
\small\emph{Institute for Nuclear Research,
National Academy of Sciences of Ukraine,} \\
\small\emph{prosp. Nauki, 47, Kiev-28, 03680, Ukraine}}

\date{\today}

\begin{document}
\begin{sloppypar}

\maketitle

\begin{abstract}
In paper SUSY-hierarchies of one-dimensional potentials with continuous
energy spectra are studied.
Use of such hierarchies for analysis of reflectionless potentials is
substantiated from the physical point of view.
An interdependence (based on Darboux transformations) between spectral
characteristics of potentials-partners is determined, an uniqueness
of its solution in result of use of boundary conditions is shown.
A rule of construction of new reflectionless potentials on the basis
of known one is corrected, its proof is proposed.
At first time a general solution for a superpotential $W_{n+m}(x)$
with an arbitrary number $n+m$ in the studied hierarchy on the basis
of only one known partial solution for the superpotential $W_{n}(x)$
with the selected number $n$ is found.

A general solution of a hierarchy of inverse power (reflectionless)
potentials is obtained.
Such a hierarchy can be interesting as an example for solution of a
known problem of search of general solutions of the hierarchies of
different types (both in standard and parasupersymmetric quantum mechanics).

A consequent statement and analysis of \emph{exactly solvable
reflectionless potentials of Gamov's type}, which at their shapes look
qualitatively like scattering potentials in two-particle description
of collisions between particles and nuclei or decay potentials in
two-particle description of decay of compound spherical nuclear systems,
are presented.
\end{abstract}



\section{Introduction
\label{sec.1}}

A widely known approach for analysis of properties of quantum systems
and processes with their participation in nuclear physics of low and
intermediate energies, in atomic and particle physics is based on
study of characteristics of potentials, described these systems.
Here, such an unusual phenomenon as \emph{resonant tunneling}, which
is shown in maximal values (close to one) of a penetrability
coefficient of the potentials (and barriers concerned with them) at
selected energy levels, has been caused a sufficient interest.
This phenomenon allows to explain an existence of the resonant values
of cross-sections of nuclear collisions, determines the resonant levels
in the energy spectra for decays of the nuclei (where the decay
probability is maximal), allows to predict half-value periods of the
nuclear decays, leads to explanation of a number of other important
effects.
Because of this, methods of description of the resonant effects are
developed intensively and are important.

A \emph{reflectionless penetration} or an \emph{absolute transparence}
of the potentials (and the barriers concerned with them) of the quantum
systems \cite{Zakhariev.1993.PHLTA,Zakhariev.1994.PEPAN},
differed from the resonant
tunneling by that it exists in the whole energy spectrum (or its large
part), where a reflection coefficient is not only minimal but equals to
zero, is more uncommon phenomenon.
This effect is studied less sufficiently and looks more unusual from a
point of view of common sense.

A lot of methods are developed for study of these and some other unusual
quantum effects (for example, \emph{reinforcement of the barrier
penetrability and breaking of tunneling symmetry in opposite directions
during the penetration of multiple particles, absolute reflection at the
above-barrier energies}). Here, we note methods of direct and inverse
problems of quantum mechanics, which are used the most widely. One note
both monographs \cite{Chadan.1977} and excellent reviews
\cite{Zakhariev.1994.PEPAN,Zakhariev.1999.PEPAN}, where both the methods
for detailed study of the properties of the reflectionless one- and
multichannel quantum systems (mainly, in the discrete energy spectra)
and simple approaches for a qualitative analysis and a clear
understanding of them are proposed.

Mainly, the methods of calculations of the spectral characteristics
(wave functions, energy spectra) in the approaches pointed out above
are approximated, while in extremely rare cases one can find types of
the potentials, where exact analytical solutions exist. These potentials,
named as \emph{exactly solvable potentials}, have caused a large
interest because of knowledge of the exact analytical form of the
spectral characteristics gives a possibility to simplify maximally
the problem of analysis of the properties of the quantum systems and,
therefore, to understand them deeper. Therefore, a lot of papers are
devoted to search of new types of the exactly solvable potentials.
However, a number of such potentials, opened by the methods of the
direct problems, are extremely small, and the found potentials are 
still far for an acceptable description of the real physical processes.
The methods of the inverse problem sufficiently expand possibilities 
to resolve this problem (in a comparison on the methods of the direct 
approach), giving enough large variety of the own exactly solvable 
models, however for construction of the exact analytical forms of 
the potentials it needs often to know full sets of the wave functions 
and the levels in the energy spectra, and this account turned out 
an enough difficult process (and can be considered as an approximated
calculating method with extremely high accuracy).

It turned out, that methods of supersymmetric quantum mechanics (SUSY QM),
developed in last two and a half decades, are sufficiently more effective 
and enough simple in solution of these tasks, than the methods of direct 
and inverse problems. With their use one can obtain new exactly solvable
potentials with their spectral characteristics.
Here, I should like to note a fine review \cite{Cooper.1995.PRPLC}, to
point out methods of Non-linear (Polynomial, $N$-fold ...)
supersymmetric quantum mechanics
\cite{Andrianov.2003.NuclPhys,Andrianov.2004.JPAGB}) developed
intensively the last decade and given own original new solutions for
the exactly solvable potentials, methods of the \emph{shape invariant
potentials} with different types of parameters transformations (for
example, see \cite{Gendenshtein.1983.JETPL,Dutt.1986.PHLTA,
Cooper.1987.PHRVA,Dutt.1988.AJPIA,Khare.1988.JPAGB,Khare.1993.JPAGB,
Barclay.1993.PHRVA,Balantekin.1997.PHRVA}), methods of description of
the \emph{self-similar potentials} studied by \emph{Shabat}
\cite{Shabat.1992.INPEE} and \emph{Spiridonov}
\cite{Spiridonov.1992.PRLTA} in details and concerned with $q$-symmetry,
methods of description of other types of the deformations of the
potentials and symmetries (for example, see
\cite{Gomez-Ullate.quant-ph/0308062}).
Also one can note a number of papers, connecting the methods of 
supersymmetry and the methods of inverse problem
\cite{Zakhariev.1994.PEPAN,Zakhariev.1999.PEPAN}
(with proposed literature list here).

As a separate subclass of the exactly solvable potentials, the methods
of SUSY QM allow to choose a set of the reflectionless potentials.
However, it failed to find the application of most of the found
exactly solvable potentials for description of the real physical
systems and processes.

It turned out, that in result of study of construction rules of
SUSY-hierarchies of the potentials with taking into account of their
energy spectra (both discrete and continuous ones) one can find the
hierarchies which have own exactly solvable general solutions, and
this gives new types of the reflectionless potentials which can be
interesting from the physical point of view.
In this paper we present a consequent receiving and analysis of the
reflectionless potentials opened in \cite{Maydanyuk.2005.APNYA}, which
in their consideration inside a spatial semi-axis have one hole and
one barrier, after which they fall down to zero monotonously in the
spatially asymptotic limit. Such potentials at their shape look
like potentials of scattering in two-particle description of
collisions between particles and nuclei or decay potentials in
two-particle description of decay of compound spherical nuclear
systems. Because of this, such potentials are named as \emph{the
potentials of Gamov's type}.
A large attention in paper is given to search of general solutions of
the (reflectionless) SUSY-hierarchies of different types.

\section{SUSY-interdependence between spectral characteristics of
potentials-partners with continuous energy spectra
\label{sec.2}}

\subsection{Darboux transformations 
\label{sec.2.1}}

In the beginning let's consider a formalism of \emph{Darboux
transformations} used widely in supersymmetric quantum mechanics
(SUSY QM) \cite{Cooper.1995.PRPLC} (see p.~275--276).
Let's consider a one-dimensional motion of a particle with mass $m$
inside a potential field $V(x)$. We introduce operators $A_{1}$ and
$A_{1}^{+}$ of the following form:
\begin{equation}
\begin{array}{ll}
  A_{1} =
  \displaystyle\frac{\hbar}{\sqrt{2m}}
  \displaystyle\frac{d}{dx}
  + W_{1}(x), &
  A_{1}^{+} =
  -\displaystyle\frac{\hbar}{\sqrt{2m}}
  \displaystyle\frac{d}{dx}
  + W_{1}(x),
\end{array}
\label{eq.2.1.1}
\end{equation}
where $W_{1}(x)$ is the function, defined on the whole spatial region
with axis $x$. We assume that this function is continuous inside the
whole region of its definition with the exception of some possible
points of discontinuity.

On the basis of the operators $A_{1}$ and $A_{1}^{+}$ we construct
Hamiltonians of the motion of the particle by two different ways
(we find the Hamiltonians of Schrodinger type only):
\begin{equation}
\begin{array}{l}
  H_{1} = A_{1}^{+} A_{1} + C_{1} =
  -\displaystyle\frac{\hbar^{2}}{2m}
  \displaystyle\frac{d^{2}}{dx^{2}}
  + V_{1}(x), \\
  H_{2} = A_{1} A_{1}^{+} + C_{1} =
  -\displaystyle\frac{\hbar^{2}}{2m}
  \displaystyle\frac{d^{2}}{dx^{2}}
  + V_{2}(x).
\end{array}
\label{eq.2.1.2}
\end{equation}
In each case the Hamiltonian (\ref{eq.2.1.2}) is expressed through

own potential $V_{1}(x)$ or $V_{2}(x)$:
\begin{equation}
\begin{array}{ll}
  V_{1}(x) =
  W_{1}^{2}(x) - \displaystyle\frac{\hbar}{\sqrt{2m}}
  \displaystyle\frac{d W_{1}(x)}{dx} + C_{1}, &
  V_{2}(x) =
  W_{1}^{2}(x) + \displaystyle\frac{\hbar}{\sqrt{2m}}
  \displaystyle\frac{d W_{1}(x)}{dx} + C_{1}.
\end{array}
\label{eq.2.1.3}
\end{equation}
We see that two potentials are expressed through one common function
$W_{1}(x)$. One can write:
\begin{equation}
  V_{2} (x) = V_{1}(x) +
    2\displaystyle\frac{\hbar}{\sqrt{2m}}
    \displaystyle\frac{d W_{1}(x)}{dx}.
\label{eq.2.1.4}
\end{equation}

Construction of the potentials $V_{1}$ and $V_{2}$ of two quantum
systems on the basis of one common function $W_{1}(x)$ establishes
interdependence between the spectral characteristics (wave functions,
energy spectra) of these systems. We shall consider this
interdependence as the interdependence, defined by Darboux
transformations.
In development of SUSY QM theory the function $W_{1}(x)$ is named as
\emph{superpotential}, while the potentials $V_{1}(x)$ and $V_{2}(x)$
are named as \emph{supersymmetric potentials-partners} (for example,
see p.~275--276 in~\cite{Cooper.1995.PRPLC}).

Note, that one constant $C_{1}$ is introduced in the definitions
(\ref{eq.2.1.2}) of the Hamiltonians of two quantum systems.
If to use $C_{1} = E^{(1)}_{0}$ (here, $E^{(1)}_{0}$ is the lowest
level of the energy spectrum of the first Hamiltonian $H_{1}$), then
we obtain the most widely used construction of two Hamiltonians
$H_{1}$ and $H_{2}$ on the basis of the operators $A_{1}$ and
$A_{1}^{+}$ (for example, see p.~287--289 in \cite{Cooper.1995.PRPLC}).
If to use $C_{1}=0$, then we obtain the construction of the
Hamiltonians of these two systems on the basis of the operators
$A_{1}$ and $A_{1}^{+}$ as in \cite{Maydanyuk.2005.APNYA} (see
p.~443, sec.~2).
At $E^{(1)}_{0} = 0$ both approaches of construction of two
Hamiltonians $H_{1}$ and $H_{2}$ are coincided:
\begin{equation}
\begin{array}{l}
  H_{1} = A_{1}^{+} A_{1} =
  -\displaystyle\frac{\hbar^{2}}{2m}
  \displaystyle\frac{d^{2}}{dx^{2}}
  + V_{1}(x), \\
  H_{2} = A_{1} A_{1}^{+} =
  -\displaystyle\frac{\hbar^{2}}{2m}
  \displaystyle\frac{d^{2}}{dx^{2}}
  + V_{2}(x).
\end{array}
\label{eq.2.1.5}
\end{equation}

\subsection{Varieties of the SUSY-hierarchies
\label{sec.2.2}}

Let's assume that there are new operators $A_{2}$ and $A_{2}^{+}$,
on the basis of which the Hamiltonian $H_{2}$ of the second system 
can be written in the form:
\begin{equation}
  H_{2} = A_{2}^{+} A_{2} + C_{2} =
  -\displaystyle\frac{\hbar^{2}}{2m}
  \displaystyle\frac{d^{2}}{dx^{2}}
  + V_{2}(x),
\label{eq.2.2.1}
\end{equation}
where a new constant $C_{2}$ is introduced.
Then on the basis of Darboux transformations one can define
SUSY-partner to the Hamiltonian $H_{2}$ --- the Hamiltonian $H_{3}$
of the third system:
\begin{equation}
  H_{3} = A_{2} A_{2}^{+} + C_{2} =
  -\displaystyle\frac{\hbar^{2}}{2m}
  \displaystyle\frac{d^{2}}{dx^{2}}
  + V_{3}(x).
\label{eq.2.2.2}
\end{equation}
By such a way one can define consecutively a new Hamiltonian $H_{n+1}$
(and a new potential $V_{n+1}$ concerned with it) with a next number
$n+1$ on the basis of the known Hamiltonian $H_{n}$ (and the potential
$V_{n}$) with the previous number $n$. Here, the following expressions
are fulfilled:
\begin{equation}
\begin{array}{l}
  H_{n} = A_{n}^{+} A_{n} + C_{n} =
  -\displaystyle\frac{\hbar^{2}}{2m}
  \displaystyle\frac{d^{2}}{dx^{2}}
  + V_{n}(x), \\
  H_{n+1} = A_{n} A_{n}^{+} + C_{n} =
  -\displaystyle\frac{\hbar^{2}}{2m}

  \displaystyle\frac{d^{2}}{dx^{2}}
  + V_{n+1}(x).
\end{array}
\label{eq.2.2.3}
\end{equation}

A sequence of such Hamiltonians (and a sequence the potentials of these
Hamiltonians) forms one SUSY-hierarchy.
If to use $C_{n}=E_{0}^{(n)}$, then we obtain the most widely used
(standard) definition of SUSY-hierarchy (for example, see p.~287--289
in~\cite{Cooper.1995.PRPLC};
here, methods of calculation of the energy spectra and wave functions
for each potential of one hierarchy, the superpotentials concerned
with them, are presented).
But on this case the lowest level of the energy spectrum of each
Hamiltonian with the next number in the hierarchy is located higher
then the lowest levels of all Hamiltonians with the previous numbers.
Usually, these hierarchies have the potentials, lower parts of the
energy spectra of which are discrete.

\emph{It turned out, that if to change values of the constants
$C_{n}$, than we obtain new types of SUSY-hierarchies with own exactly
solvable potentials, which can be differed from the potentials of the
hierarchy, defined by the way pointed out above.}
Here, if to use $C_{n}=0$, then we obtain the hierarchy of the
potentials, studied in~\cite{Maydanyuk.2005.APNYA}.
In such hierarchy the lowest level of the Hamiltonian with arbitrary
number coincides with the lowest levels of the Hamiltonians with the
other numbers.
In particular, if to select $C_{n}=E_{0}^{(n)}=0$, then these
hierarchies turn out to be convenient enough for connection together
of the Hamiltonians with the continuous energy spectra completely,
and their potentials turn out to be useful for study of real processes
of particles and nuclei scattering (because of the scattering potentials 
tend to zero in asymptotic spatial regions).
\emph{Therefore, the hierarchies of such a type are more useful from
the physical point of view in study of collisions between particles
and nuclei in a comparison on the most widely used type of the
hierarchy presented, for example, in~\cite{Cooper.1995.PRPLC} 
(see p.~287-289), and their use is substantiated physically for study of 
the reflectionless potentials.}

Because of this, we shall consider further just this case.

\subsection{Interdependence between wave functions of the
potentials-partners
\label{sec.2.3}}

Let's consider two quantum systems with the continuous energy spectra.
Then Darboux transformations allow to establish an interdependence
between the wave functions of these systems. In accordance with
(\ref{eq.2.1.5}), we write:
\begin{equation}
\begin{array}{l}
  H_{1} \varphi^{(1)}_{k} =
  A_{1}^{+} A_{1} \varphi^{(1)}_{k} =
  E^{(1)}_{k} \varphi^{(1)}_{k}, \\
  H_{2} \varphi^{(2)}_{k^{\prime}} =
  A_{1} A_{1}^{+} \varphi^{(2)}_{k^{\prime}} =
  E^{(2)}_{k^{\prime}} \varphi^{(2)}_{k^{\prime}},
\end{array}
\label{eq.2.3.1}
\end{equation}
where $E^{(1)}_{k}$ and $E^{(2)}_{k^{\prime}}$ are the energy levels of
these systems,
$\varphi^{(1)}_{k}(x)$ and $\varphi^{(2)}_{k^{\prime}}(x)$ are
the wave functions (eigen-functions) corresponding to these levels,
$k = \displaystyle\frac{1}{\hbar}\sqrt{2mE^{(1)}_{k}}$ and
$k^{\prime} =
\displaystyle\frac{1}{\hbar}\sqrt{2mE^{(2)}_{k^{\prime}}}$ are wave
vectors corresponding to the levels $E^{(1)}_{k}$ and
$E^{(2)}_{k^{\prime}}$.
From (\ref{eq.2.3.1}) we obtain:

\begin{equation}
  H_{2} (A_{1} \varphi^{(1)}_{k}) =
  A_{1} A_{1}^{+} (A_{1} \varphi^{(1)}_{k}) =
  A_{1} (A_{1}^{+} A_{1} \varphi^{(1)}_{k}) =
  A_{1} (E^{(1)}_{k} \varphi^{(1)}_{k}) =
  E^{(1)}_{k} (A_{1} \varphi^{(1)}_{k}).
\label{eq.2.3.2}
\end{equation}
We see, that the function $f(x) = A_{1} \varphi^{(1)}_{k} (x)$ is
the eigen-function of the operator $\hat{H}_{2}$ to a constant factor,
i.~e. it represents the wave function $\varphi^{(2)}_{k^{\prime}} (x)$
of the Hamiltonian $H_{2}$.
The energy level $E^{(1)}_{k}$ must be the eigen-value of this
operator exactly, i.~e. it represents the energy level
$E^{(2)}_{k^{\prime}}$ of this Hamiltonian. Here, new wave function
and energy level have the same index ${k^{\prime}}$. One can write:
\begin{equation}
\begin{array}{lcr}
  A_{1} \varphi^{(1)}_{k} (x) =
    N_{2} \varphi^{(2)}_{k^{\prime}} (x), &
  E^{(1)}_{k} = E^{(2)}_{k^{\prime}}, &
  N_{2} = const.
\end{array}
\label{eq.2.3.3}
\end{equation}

Taking into account (\ref{eq.2.3.1}), we write:
\begin{equation}
  H_{1} (A_{1}^{+} \varphi^{(2)}_{k^{\prime}}) =
  A_{1}^{+} A_{1} (A_{1}^{+} \varphi^{(2)}_{k^{\prime}}) =
  A_{1}^{+} (A_{1} A_{1}^{+} \varphi^{(2)}_{k^{\prime}}) =
  A_{1}^{+} (E^{(2)}_{k^{\prime}} \varphi^{(2)}_{k^{\prime}}) =
  E^{(2)}_{k^{\prime}} (A_{1}^{+} \varphi^{(2)}_{k^{\prime}})
\label{eq.2.3.4}
\end{equation}
and obtain:
\begin{equation}
\begin{array}{lcr}
  A_{1}^{+} \varphi^{(2)}_{k^{\prime}} (x) =
    N_{1} \varphi^{(1)}_{k} (x), &
  E^{(2)}_{k^{\prime}} = E^{(1)}_{k}, &
  N_{1} = const.
\end{array}
\label{eq.2.3.5}
\end{equation}

Therefore, we find the following interdependences between the wave
functions and the energy levels for two systems SUSY-partners in
regions of the continuous energy spectra:
\begin{equation}
\begin{array}{lcr}
  \varphi^{(1)}_{k} (x) =
    \displaystyle\frac{1}{N_{1}}
    A_{1}^{+} \varphi^{(2)}_{k^{\prime}} (x), &
  \varphi^{(2)}_{k^{\prime}} (x) =
    \displaystyle\frac{1}{N_{2}}
    A_{1} \varphi^{(1)}_{k} (x), &
  E^{(1)}_{k} = E^{(2)}_{k^{\prime}}.
\end{array}
\label{eq.2.3.6}
\end{equation}

\vspace{3mm}
{\bf Conclusion:}
\emph{
Expressions (\ref{eq.2.3.6}) prove, that the hierarchy, defined at
$C_{n}=0$, establishes the interdependence (on the basis of Darboux
transformations with the operators $A_{1}$ and $A_{1}^{+}$ of the
form (\ref{eq.2.1.1})) between the wave functions and the levels of
the continuous energy spectra of the Hamiltonians-partners.}

\vspace{2mm}
{\bf Consequence 1:}
\emph{
One can calculate coefficients $N_{1}$ and $N_{2}$ uniquely from
normalization conditions of the wave functions (for the continuous
energy spectrum, with taking into account of boundary conditions) for
each system.
A selection and use of the boundary conditions make the solution of the
interdependence between the wave functions of the potentials-partners
with the continuous energy spectra as uniqueness
(the interdependence (\ref{eq.2.3.6}) between the wave functions
is corrected in a comparison on Exp.~(14) in~\cite{Maydanyuk.2005.APNYA}).}

\vspace{2mm}
{\bf Consequence 2:}
\emph{
In accordance with (\ref{eq.2.3.6}), the energy levels of the 
Hamiltonians-partners with the continuous energy spectra of the 
hierarchy, defined at $C_{n}=0$, coincide between each other
(as for the Hamiltonians of the hierarchy with the discrete
energy spectra with a possible exception of the lowest levels;
the SUSY-interdependence between the energy levels is corrected in a
comparison on Exp.~(14) in \cite{Maydanyuk.2005.APNYA}).}

\vspace{2mm}
{\bf Consequence 3:}
\emph{
Uniquely interdependence between the wave functions and the energy
levels of the systems-partners of the hierarchy, defined at $C_{n}=0$,
establishes an uniquely interdependence between other characteristics
of these systems, calculation of which is based on the wave functions
and the energy spectra.}

\subsection{Interdependence between coefficients of penetrability and
reflection
\label{sec.2.4}}

We shall consider a case when the superpotential $W_{1}(x)$ and the
potentials $V_{1}(x)$ and $V_{2}(x)$ of two studied systems are finite
in the whole spatial region of their definition. At $x \to \pm\infty$
one can write:
\begin{equation}
  W_{1} (x \to \pm\infty) = W_{\pm}
\label{eq.2.4.1}
\end{equation}
and
\begin{equation}
  V_{1} (x \to \pm\infty) = V_{2} (x \to \pm\infty) = W^{2}_{\pm}.
\label{eq.2.4.2}
\end{equation}
On this case one can determine amplitudes of transmission and
reflection for the wave functions of these systems, and Darboux
transformations establish an unique dependence between them (for
example, see \cite{Cooper.1995.PRPLC}, p.~278--279).

Let's consider a plane wave $e^{ik_{-}x}$ which propagates in a
positive direction of the axis $x$ inside fields of two potentials
$V_{1}(x)$ and $V_{2}(x)$ (we assume that this plane wave has the
same wave vector $k_{-}$ for the fields of two potentials).
Then in spatial asymptotic regions we obtain the transmitted waves
$T_{1}(k_{-}, k_{+}) e^{ik_{+}x}$ and $T_{2}(k_{-}, k_{+}) e^{ik_{+}x}$,
and the reflected waves $R_{1}(k_{-}) e^{-ik_{-}x}$ and $R_{2}(k_{-})
e^{-ik_{-}x}$ also. For the wave function one can write:
\begin{equation}
\begin{array}{ll}
  \varphi^{(1)}(k, x \to -\infty) \to
    \bar{N}_{1} (e^{ik_{-}x} + R_{1} e^{-ik_{-}x}), &
  \varphi^{(1)}(k, x \to +\infty) \to
    \bar{N}_{1} T_{1} e^{ik_{+}x}, \\
  \varphi^{(2)}(k, x \to -\infty) \to
    \bar{N}_{2} (e^{ik_{-}x} + R_{2} e^{-ik_{-}x}), &
  \varphi^{(2)}(k, x \to +\infty) \to
    \bar{N}_{2} T_{2} e^{ik_{+}x},
\end{array}
\label{eq.2.4.3}
\end{equation}
where $k$, $k_{-}$ and $k_{+}$ are defined by such a way:
\begin{equation}
\begin{array}{lcr}
  k = \displaystyle\frac{1}{\hbar} \sqrt{2mE}, &
  k_{-} = \displaystyle\frac{1}{\hbar} \sqrt{2m(E-W^{2}_{-})}, &
  k_{+} = \displaystyle\frac{1}{\hbar} \sqrt{2m(E-W^{2}_{+})},
\end{array}
\label{eq.2.4.4}
\end{equation}
and coefficients $\bar{N}_{1}$ and $\bar{N}_{2}$ can be obtained
from the normalization conditions with taking into account of the
potentials forms and the boundary conditions.

Using the asymptotic expressions (\ref{eq.2.4.3}) for the wave
functions, taking into account the interdependence (\ref{eq.2.3.6})
between them and the definitions (\ref{eq.2.1.1}) for the operators
$A_{1}$ and $A_{1}^{+}$, we obtain:
\begin{equation}
\begin{array}{l}
  \bar{N}_{1} \biggl( e^{ik_{-}x} + R_{1} e^{-ik_{-}x} \biggr) =
    \displaystyle\frac{\bar{N}_{2}}{N_{1}}
    \biggl[
    \biggl(W_{-} - \displaystyle\frac{ik_{-}\hbar}{\sqrt{2m}} \biggr)
    e^{ik_{-}x} +
    \biggl(W_{-} + \displaystyle\frac{ik_{-}\hbar}{\sqrt{2m}} \biggr)
    R_{2} e^{-ik_{-}x} \biggr], \\
  \bar{N}_{1} T_{1} e^{ik_{+}x} =
    \displaystyle\frac{\bar{N}_{2}}{N_{1}} T_{2}
    \biggl(W_{+} - \displaystyle\frac{ik_{+}\hbar}{\sqrt{2m}} \biggr)

     e^{ik_{+}x}.
\end{array}
\label{eq.2.4.5}
\end{equation}
These expressions are found only for the asymptotic spatial regions
for the systems-partners of the hierarchy, defined at $C_{n}=0$.
They are fulfilled only in a case when items with the same exponents
are equal between each other. We obtain:
\begin{equation}
  \displaystyle\frac{\bar{N}_{2}}{N_{1}} =
    \displaystyle\frac
    {\bar{N}_{1}}
    {W_{-} - \displaystyle\frac{ik_{-}\hbar}{\sqrt{2m}}}
\label{eq.2.4.6}
\end{equation}
and 
\begin{equation}
\begin{array}{lr}
  R_{1}(k_{-}) =
    R_{2}(k_{-}) \displaystyle\frac
    {W_{-} + \displaystyle\frac{ik_{-}\hbar}{\sqrt{2m}}}
    {W_{-} - \displaystyle\frac{ik_{-}\hbar}{\sqrt{2m}}}, &
  T_{1}(k_{-}, k_{+}) =
    T_{2}(k_{-}, k_{+}) \displaystyle\frac
    {W_{+} - \displaystyle\frac{ik_{+}\hbar}{\sqrt{2m}}}
    {W_{-} - \displaystyle\frac{ik_{-}\hbar}{\sqrt{2m}}}.
\end{array}
\label{eq.2.4.7}
\end{equation}

Expressions (\ref{eq.2.4.7}) establish the unique interdependence
between the amplitudes of the transmission and the reflection for
two quantum systems (it agrees with \emph{Consequence 3}, pointed out
in the previous section).
One can define coefficients of penetrability and reflection of the
potentials $V_{1}(x)$ and $V_{2}(x)$ as squares of modules of the
amplitudes of the transmission and the reflection.
We see that all these coefficients are not depended on the normalized
coefficients $N_{1}$, $N_{2}$, $\bar{N}_{1}$, $\bar{N}_{2}$ and they
are defined relatively one selected energy level, along which the
propagation of the wave is studied.

We note, that the expressions of the interdependences between
the amplitudes (and the coefficients) of the transmission and the 
reflection are known mainly for the systems with the energy spectra, 
containing low discrete levels. We repeat their receiving for 
hierarchies of potentials (i.~e. a lot of potentials) with the 
continuous energy spectra completely.

\section{Search of a hierarchy of reflectionless potentials 
\label{sec.3}}

Let's consider a quantum system with a potential, which has zero
coefficient of reflection. Then a wave function of such system has
an amplitude of the reflection, which equals to zero also.
We shall name these quantum systems and their potentials as
\emph{reflectionless} or \emph{absolutely transparent}.
Then from (\ref{eq.2.4.7}) one can see that a potential SUSY-partner
(and a system corresponding to it) for the reflectionless potential
is reflectionless also.
Using this simple idea and knowing a form of only one reflectionless
potential, one can construct a number of new (early unknown) exactly
solvable reflectionless potentials.

\subsection{A recurrent method of construction of new reflectionless
potentials
\label{sec.3.1}}

Now we formulate the following

\vspace{3mm}
{\bf Rule:}
\emph{
two potentials $V_{1}(x)$ and $V_{2}(x)$ are reflectionless, if the
following conditions are fulfilled:
}
\begin{itemize}

\item
\vspace{-3mm}
\emph{these potentials are potentials SUSY-partners;}

\vspace{-3mm}
\item
\emph{one potential is reflectionless;}

\vspace{-3mm}
\item
\emph{asymptotic expressions for the wave functions for both potentials
have forms
\begin{equation}
\begin{array}{l}
  \varphi^{(1,2)}(k, x \to -\infty) \to e^{ikx} + R_{1,2} e^{-ikx}, \\
  \varphi^{(1,2)}(k, x \to +\infty) \to T_{1,2} e^{ik'x},
\end{array}
\label{eq.3.1.1}
\end{equation}
}
\end{itemize}

This \emph{Rule} was proposed in \cite{Maydanyuk.2005.APNYA}
(see p.~447, sec.~3) and we correct its formulation in this paper.
Its proof consists in fulfillment of the condition (\ref{eq.2.4.7}) for
the potentials-partners with continuous energy spectra and its
application to the reflectionless potentials.
Here, we consider only such a hierarchy, where interdependences
between spectral characteristics of the potentials-partners are based
on Darboux transformations.
It does not take into account operator transformations of
\emph{non-linear supersymmetry}
\cite{Andrianov.2003.NuclPhys,Andrianov.2004.JPAGB},
which we do not study in this paper.
However, this idea is applied in that case also.

A constant potential does not give the reflection for a plane wave
in its propagation inside this potential. Because of this, one can
consider the constant potential as the refectionless one.
Therefore, we conclude:

\vspace{3mm}
{\bf Consequence 1}:
\emph{
If a potential belongs to SUSY-hierarchy, which has the constant
potential, and its wave function in asymptotic regions can be
presented in the form (\ref{eq.3.1.1}), then this potential is
reflectionless.
}

\vspace{3mm}
In \cite{Maydanyuk.2005.APNYA} (see p.~447--448, sec.~3) there is
the following

\vspace{3mm}
{\bf Supposition 1}:
\emph{
A potential is reflectionless only in such a case, if it belongs to
SUSY-hierarchy, which has the constant potential, and its wave
function in the asymptotic regions can be written as (\ref{eq.3.1.1}).
}

\vspace{3mm}
One can formulate this supposition by another way 

\vspace{3mm}
{\bf Supposition 2}:
\emph{
a reason of the reflectionless property of any potential consists in its 
SUSY-interdependence with the constant potential.}

\vspace{3mm}
A proof of the \emph{Rule}, formulated above, is a needed condition
(proof of necessity) of fulfillment of the \emph{Suppositions 1} and
\emph{2}.
However, a proof of sufficiency of these suppositions must to take
into account all possible types of SUSY-interdependences between the

operators $A$ and $A^{+}$ of algebra of supersymmetry (Darboux
transformations, operator transformations of non-linear supersymmetry
are partial cases of this algebra) and we do not know it else.

On the basis of the \emph{Consequence 1}, formulated above, one can
construct by recurrent way a hierarchy of the reflectionless potentials, 
if to use the constant potential as the first (started) potential.

Here, if to use the constant potential of a form:
\begin{equation}
  V_{1}(x) = A^{2} = \mbox{const}
\label{eq.3.1.2}
\end{equation}
(we shall take into account cases $A^{2} \le 0$ and $A^{2} \ge 0$ at
$A^{2} \in \mathcal{R}$),
then one can find for it a general form of a superpotential (see (33)
and (37) in p.~449--450 in~\cite{Maydanyuk.2005.APNYA}), which
connects this potential with all possible types of the potentials-partners.
After obtaining the superpotential, one can find all possible
solutions for the potentials-partners to the constant potential.
In accordance with \cite{Maydanyuk.2005.APNYA} (see (34) and (38) in
p.~450), this approach gives only five partial solutions for the
potential-partner ($\alpha = \hbar / \sqrt{2m}$):
\begin{equation}
\begin{array}{l}
  V_{2}(x) =
  \left\{
  \begin{array}{cl}

    A^{2}, & W^{2}_{1}(x) = A^{2}; \\
    \displaystyle\frac{2\alpha^{2}}{(x+x_{0})^{2}},
    & W^{2}_{1}(x) \ne A = 0 \mbox{ and } x+x_{0} \ne 0; \\
    |A|^{2} \left(1 + \displaystyle\frac{2}{\sinh^{2}
      \bigl(|A|(x+x_{0}) / \alpha \bigr)} \right),
    & A^{2} > 0 \mbox{ and } |W_{1}(x)| > |A|; \\
    |A|^{2} \left(1 - \displaystyle\frac{2}{\cosh^{2}
    \bigl(|A|(x+x_{0}) / \alpha \bigr)} \right),
    & A^{2} > 0 \mbox{ and } |W_{1}(x)| < |A|; \\
    |A|^{2} \left(\displaystyle\frac{2}{\cos^{2}
    \bigl(|A|(x+x_{0}) / \alpha \bigr)} - 1 \right),
    & A^{2} < 0 \mbox{ and } W_{1}^{2}(x) \ne A^{2}
  \end{array}
  \right.
\end{array}
\label{eq.3.1.3}
\end{equation}
and there are no other solutions.
Further, with use of the \emph{Rule}, formulated above, from the found
five solutions of the system (\ref{eq.3.1.3}) one can select the
reflectionless potentials, which form the general solution for the
reflectionless potential-partner to the constant potential.
Here, in accordance with the third point of the \emph{Rule}, the
fifth solution of the system (\ref{eq.3.1.3}) cannot be reflectionless
potential, because of its wave function cannot be presented in the
form (\ref{eq.3.1.1}).
An effectiveness of introduction of the third point into the
\emph{Rule} (unlike the approach in \cite{Cooper.1995.PRPLC} 
and other papers, known to us and concerned with the reflectionless 
potentials) is shown in this.

Further, using the found reflectionless potentials of the system
(\ref{eq.3.1.3}), one can obtain on the basis of the
\emph{Consequence 1} new reflectionless potentials with the next
numbers and, therefore, construct sequentially by recurrent way
the hierarchy of the reflectionless potentials.

\subsection{A general solution for a superpotential with an
arbitrary number
\label{sec.3.2}}

For finding the superpotential $W_{n}(x)$ with the selected number $n$
in the hierarchy, it needs to solve the \emph{Ricatti equation}. In
accordance with \cite{Tihonov.1998} (see p.~29), in a general case
this equation cannot be integrated. However, it has one property:
\emph{if we know one a partial solution of this equation, then one
can reduce this equation to the Bernoulli equation and find its
general solution}.

As we found above, one can obtain the general solution for the
superpotential $W_{1}(x)$, connected the constant potential with its
potential-partner.

Now we assume, that we know a partial solution for a superpotential
$\tilde{W}_{n}(x)$ with a number $n$ in the hierarchy. Then from the
following transformations ($\alpha = \hbar / \sqrt{2m}$)
\[
\begin{array}{l}
  \left.
  \begin{array}{l}
    V_{n+1}(x) =
    W_{n}^{2}(x) + \alpha \displaystyle\frac{d W_{n}(x)}{dx}; \\
    V_{n+1}(x) =
    W_{n+1}^{2}(x) - \alpha \displaystyle\frac{d W_{n+1}(x)}{dx}
  \end{array}
  \right \}

  \Longrightarrow
\end{array}
\]
\begin{equation}
\begin{array}{ll}
  \Longrightarrow &
  W_{n+1}^{2}(x) - \alpha \displaystyle\frac{d W_{n+1}(x)}{dx} =
  W_{n}^{2}(x) + \alpha \displaystyle\frac{d W_{n}(x)}{dx}.
\end{array}
\label{eq.3.2.1}
\end{equation}
one can see, that a partial solution (early unknown) for a superpotential
$\tilde{W}_{n+1}(x)$ with a next number $n+1$ in the hierarchy can be
presented in the form:
\begin{equation}
  \tilde{W}_{n+1}(x) = -\tilde{W}_{n}(x).
\label{eq.3.2.2}
\end{equation}
Therefore, we have proved, that in one SUSY-hierarchy, which is
defined at $C_{n}=0$, there are two partial solutions for the
superpotentials with the neighboring numbers, which are connected
by (\ref{eq.3.2.2}).

Further, knowing the parial solution for the superpotential
$\tilde{W}_{n+1}(x)$ with the number $n+1$, by such a way one can find
a partial solution for a superpotential $\tilde{W}_{n+2}(x)$ with a
next number $n+2$, which coincides with the superpotential
$\tilde{W}_{n}(x)$.
Continuing this logics further, we obtain recurrently an expression
for a partial solution for a superpotential $\tilde{W}_{n+m}(x)$ with
an arbitrary number $n+m$:
\begin{equation}
  \tilde{W}_{n+m}(x) = (-1)^{m} \tilde{W}_{n}(x).
\label{eq.3.2.3}
\end{equation}
Here, we have proved, that in one SUSY-hierarchy, defined at $C_{n}=0$,
for the known partial solution for the superpotential
$\tilde{W}_{n}(x)$ with the selected number $n$ there is the partial
solution for the superpotential $\tilde{W}_{n+m}(x)$ with the
arbitrary number $n+m$, and these superpotentials are connected by
(\ref{eq.3.2.2}) (it is found for the first time).

\vspace{3mm}
{\bf Conclusion}:
\begin{itemize}

\item
\emph{
Knowing the partial solution for the superpotential $\tilde{W}_{n}(x)$
with the selected number $n$ in the hierarchy, one can find the partial
solution for the superpotential $\tilde{W}_{n+1}(x)$ with the next
number $n+1$ in the hierarchy by use of (\ref{eq.3.2.2})}.

\item
\emph{
Knowing the partial solution for the superpotential $\tilde{W}_{n}(x)$
with the selected number $n$ in the hierarchy, one can find the partial
solution for the superpotential $\tilde{W}_{n+m}(x)$ with the arbitrary
number $n+m$ in the hierarchy by use of (\ref{eq.3.2.3}) (it is found
for the first time)}.
\end{itemize}

\vspace{3mm}
Further, knowing the partial solution for the superpotential
$\tilde{W}_{n+m}(x)$, we shall find its general solution. Let's solve
the Ricatti equation (\ref{eq.3.2.1}), rewriting it in the form:
\begin{equation}
  \displaystyle\frac{d W_{n+m}(x)}{dx} -
  \displaystyle\frac{1}{\alpha} W_{n+m}^{2}(x) =
  - \displaystyle\frac{F(x)}{\alpha}.
\label{eq.3.2.4}
\end{equation}

A change
\begin{equation}
  y(x) = \displaystyle\frac{1} {W_{n+m}(x) - \tilde{W}_{n+m}(x)}
\label{eq.3.2.5}
\end{equation}
transforms the equation (\ref{eq.3.2.4}) into the form:
\begin{equation}
  \displaystyle\frac{dy(x)}{dx} +
  \displaystyle\frac{2\tilde{W}_{n+m}(x)}{\alpha} y(x) =
  - \displaystyle\frac{1}{\alpha}.
\label{eq.3.2.6}
\end{equation}

A general solution of this equation is:
\begin{equation}
  y(x) =
  e^{-\displaystyle\int \displaystyle\frac{2\tilde{W}_{n+m}(x)}{\alpha} dx}
  \Biggl( C_{1} -
  \displaystyle\frac{1}{\alpha}
  \displaystyle\int
  e^{\displaystyle\int
     \displaystyle\frac{2\tilde{W}_{n+m}(x)}{\alpha} dx} dx \Biggl).
\label{eq.3.2.7}
\end{equation}
where $C_{1}$ is a constant of integration.
We write this equation in the old variables:
\begin{equation}
  W_{n+m}(x) =
  \displaystyle\frac{1}{y(x)} + \tilde{W}_{n+m}(x) =
  \displaystyle\frac{\mathstrut
    \alpha
    e^{\textstyle\frac{2}{\alpha}
    \displaystyle\int \tilde{W}_{n+m}(x) dx}}
  {\alpha C_{1} -
     \displaystyle\int
     e^{\textstyle\frac{2}{\alpha}
        \displaystyle\int \tilde{W}_{n+m}(x) dx}
     dx}
  + \tilde{W}_{n+m}(x).
\label{eq.3.2.8}
\end{equation}
Rewrite this solution through the given superpotential
$\tilde{W}_{n}(x)$ with the selected number $n$:
\begin{equation}
  W_{n+m}(x) =
  \displaystyle\frac{\mathstrut
    \alpha
    e^{(-1)^{m} \textstyle\frac{2}{\alpha}
    \displaystyle\int \tilde{W}_{n}(x) dx}}
  {\alpha C_{1} -
     \displaystyle\int
     e^{(-1)^{m} \textstyle\frac{2}{\alpha}
        \displaystyle\int \tilde{W}_{n}(x) dx}
     dx}
  + (-1)^{m} \tilde{W}_{n}(x).
\label{eq.3.2.9}
\end{equation}

Therefore, we have found the general solution for the superpotential
$W_{n+m}(x)$ with the arbitrary number $n+m$ in the hierarchy on the
basis of the known partial solution for the superpotential
$\tilde{W}_{n}(x)$ with the selected number $n$ (it is found for the
first time).
Obtaining the general solution for the superpotential with the
selected number, one can calculate further (uniquely) a general
solution for the potentials-partners, connected with this
superpotential. So, one can determine a general solution for
SUSY-hierarchy, which is defined at $C_{n}=0$ and when only one
partial solution for the superpotential with the selected number is
known for it.

If to use the constant potential as the first potential, then the
method described above allows to construct a general form of the
hierarchy of the reflectionless potentials, defined at $C_{n}=0$.
We note, that in~\cite{Maydanyuk.2005.APNYA} a recurrent approach
for  construction of the general solution for the hierarchy of the
reflectionless potentials was proposed only.

The formula for the superpotential has the integral form and contains
a dependence on one arbitrary constant of integration $C_{1}$.
Changing this constant, one can change the shape of the
potentials-partners, connected with this superpotential, without
replacement of locations of the energy levels. Therefore, we have
obtained a set of \emph{isospectral potentials} with the same
numbers in the hierarchy.

\section{Construction of new exactly solvable reflectionless potentials
\label{sec.4}}

\subsection{Hierarchy of inverse power potentials: a general solution 
\label{sec.4.1}}

Let's consider the second solution of the system (\ref{eq.3.1.3}).
As it is shown in \cite{Maydanyuk.2005.APNYA} (see p.~452--454,
sec.~5.1.1) and in \cite{Maydanyuk.2004.PAST.refl}, this potential can
be reflectionless and it has been causing an interest for study of
properties of the reflectionless potentials.

We shall construct a hierarchy, which includes the potentials of such
inverse power type only. In a general case one can write the potential
$V_{n} (x)$ of this hierarchy in a form:
\begin{equation}
  V_{n} (x) = \left\{
  \begin{array}{ll}
    \gamma_{n} \displaystyle\frac{\alpha^{2}}{(x-x_{0})^{2}}, &
      \mbox{at } x<0; \\
    \gamma_{n} \displaystyle\frac{\alpha^{2}}{(x+x_{0})^{2}}, &
      \mbox{at } x>0,
  \end{array} \right.
\label{eq.4.1.1}
\end{equation}
where $\gamma_{n} = const$.

At first, we shall find a superpotential $W_{n}(x)$, concerned with
this potential, supposing that this potential is known. One can
determine the superpotential from the following equation:
\begin{equation}
  \alpha \displaystyle\frac{d W_{n}(x)}{dx} =
  W_{n}^{2}(x) - V_{n}(x).
\label{eq.4.1.2}
\end{equation}
In \cite{Maydanyuk.2005.APNYA} a \emph{partial solution}
$\tilde{W}_{n}(x)$ of this equation, which determines a
potential-partner $V_{n+1}(x)$ of the inverse power type
(\ref{eq.4.1.1}) with a next number $n+1$ in the hierarchy, was found
(we write it for $x>0$):
\begin{equation}
\begin{array}{ll}
  \tilde{W}_{n}(x) = -\displaystyle\frac{\beta_{n}}{x+x_{0}}, &
  \beta_{n} = const.
\end{array}
\label{eq.4.1.3}
\end{equation}

As we see, this solution has the inverse power form (and it can be
tested by a simple substitution into the equation (\ref{eq.4.1.2})
with taking into account (\ref{eq.4.1.1})).
Let's find a general solution of the superpotential $W_{n}(x)$,
defined by the equation (\ref{eq.4.1.2}). We see, that this equation
(\ref{eq.4.1.2}) looks like the equation (\ref{eq.3.2.3}) (where it
needs to replace the indexes $n+1 \to n$ and the functions
$F(x) \to V_{n}(x)$). Therefore, the general solution for the
superpotential $W_{n}(x)$ can be written in the form (\ref{eq.3.2.8}),
where one need to use the solution (\ref{eq.4.1.3}) as the partial
solution. We obtain:
\begin{equation}
  W_{n}(x) =
    \left \{
    \begin{array}{cl}
      \displaystyle\frac{2\beta_{n} - \alpha}
        {C_{1}(2\beta_{n} - \alpha)
        \bar{x}^{2\beta_{n} / \alpha} + \bar{x}} -
        \displaystyle\frac{\beta_{n}}{\bar{x}}, &
        \mbox{at } 2\beta_{n} \ne \alpha; \\
      \displaystyle\frac{\alpha}{\bar{x}}
        \biggl(
        \displaystyle\frac{1}{\alpha C_{1} - \log{\bar{x}}} -
        \displaystyle\frac{1}{2} \biggr), &
        \mbox{at } 2\beta_{n} = \alpha,
    \end{array} \right.
\label{eq.4.1.4}
\end{equation}
where $\bar{x} = x-x_{0}$ at $x<0$ and $\bar{x} = x+x_{0}$ at $x>0$.
The potential-partner $V_{n+1}(x)$ to the potential (\ref{eq.4.1.1}),
defined by the superpotential (\ref{eq.4.1.4}), is not inverse power
one in a general case and its shape is deformed at a change of the
parameter $C_{1}$ (without replacement of locations of the energy
levels in spectra).

If to require that the new potential-partner $V_{n+1}(x)$ must be the
inverse power type (\ref{eq.4.1.1}) only, then we obtain uniquely a
requirement (we consider a case $2\beta_{n} \ne \alpha$):
\begin{equation}
  C_{1} = 0,
\label{eq.4.1.5}
\end{equation}
which reduces the general solution (\ref{eq.4.1.4}) into the form
(\ref{eq.4.1.3}).
Therefore, we have proved the following 

\vspace{3mm}
{\bf Property 1:}
\emph{
the solution (\ref{eq.4.1.3}) for the superpotential $W_{n}(x)$ is its
general solution, which defines the general form of the potentials of
the inverse power type (\ref{eq.4.1.5}) in one hierarchy (it is found
at the first time).}

\vspace{3mm}
Substituting the solution (\ref{eq.4.1.3}) as the general solution into
(\ref{eq.4.1.2}) and taking into account (\ref{eq.4.1.1}), we obtain
$\beta_{n}$:
\begin{equation}
\begin{array}{ll}
  W_{n}(x) =
    \displaystyle\frac{ -\alpha (1 \pm \sqrt{4\gamma_{n}+1})}
    {2(x+x_{0})}, &
  \beta_{n} =
    \displaystyle\frac{\alpha}{2}
    \biggl(1 \pm \sqrt{4\gamma_{n}+1} \biggr).
\end{array}
\label{eq.4.1.6}
\end{equation}
From here we see, that there is the following 

\vspace{3mm}
{\bf Property 2 (uniqueness of construction of the hierarchy of the
inverse power potentials):}
\emph{the general solution of the superpotential $W_{n}(x)$, which
determines uniquely the potential-partner $V_{n+1}(x)$ with the next
number $n+1$ in the hierarchy, is determined uniquely by the given
potential $V_{n}(x)$ with the previous number $n$ (i.~e. its
coefficients $\gamma_{n}$) (it is found for the first time)}.

\vspace{3mm}
From (\ref{eq.4.1.6}) we obtain:
\begin{equation}
\begin{array}{lcl}
  V_{n+1}(x) & = &
  W_{n}^{2}(x) + \alpha \displaystyle\frac{d W_{n}(x)}{dx} =
  \biggl(1 + \gamma_{n} \pm \sqrt{4\gamma_{n}+1} \biggr)
  \displaystyle\frac{\alpha^{2}}{(x+x_{0})^{2}}.
\end{array}
\label{eq.4.1.7}
\end{equation}

We have found the general form of the potential in the hierarchy of
the inverse power potentials (unlike \cite{Maydanyuk.2005.APNYA}, see
p.~452--455, sec.~5.1.2, where the partial solution was found only).
We write it in such a form:
\begin{equation}
\begin{array}{ll}
  V_{n}(x) = \left\{
  \begin{array}{ll}
    \gamma_{n} \displaystyle\frac{\alpha^{2}}{(x-x_{0})^{2}}, &
      \mbox{at } x<0; \\
    \gamma_{n} \displaystyle\frac{\alpha^{2}}{(x+x_{0})^{2}}, &
      \mbox{at } x>0; \\
  \end{array} \right. &
  \gamma_{n \pm 1} = 1 + \gamma_{n} \pm \sqrt{4\gamma_{n}+1}.
\end{array}
\label{eq.4.1.8}
\end{equation}

If to use the constant potential as the first (starting) potential
$V_{1}(x)$ in this hierarchy (and, therefore, we use $\gamma_{1}=0$),
then in accordance with the \emph{Consequence~1}, formulated and proved
in sec.~3.1, the hierarchy of the inverse power potentials
(\ref{eq.4.1.8}) becomes the \emph{hierarchy of the reflectionless
inverse power potentials} and the solution (\ref{eq.4.1.8}) becomes
the general solution of the potential with the arbitrary number $n$
in this hierarchy.

Using the recurrent expressions (\ref{eq.4.1.8}), one can calculate
values of the coefficients $\gamma_{n}$ with next numbers in this
hierarchy. We write first some values
\begin{equation}
  \gamma_{n} = 0, 2, 6, 12, 20, 30, 42...
\label{eq.4.1.9}
\end{equation}
Note, when the hierarchy of the inverse power potentials becomes

reflectionless, then the coefficients $\gamma$ become natural numbers.

So, we have shown, that \emph{knowing only one partial solution for
the superpotential with the selected number in the hierarchy, defined
at $C_{n}=0$, one can determine uniquely the general solution for all
potentials in this hierarchy (with taking into account of needed form
of these potentials), i.~e. one can find the general solution for the
hierarchy of the needed type.
Such a hierarchy can be interesting as an example of solution of a
known problem of a search of general solutions of hierarchies, defined
at the different values of $C_{n}$.}
Here, in \cite{Cooper.1995.PRPLC} (see p.~374, sec.~13.1) a problem
of search of a general form of an interdependence between
superpotentials with neighboring numbers in hierarchies of
parasupersymmetric quantum mechanics, defined at $C_{n}=E_{n}^{(1)}$,
is pointed out.
In \cite{Rubakov.1993.MPLAE} non-linear equations, connected variables
$W_{n} \pm W_{n-1}$ together, are obtained and are solved easily.
However, it is not clear, how to obtain a form of the superpotential
with a next number on the basis of the known superpotential with a
previous number.

Note, that the hierarchy of the inverse power reflectionless potentials
represents a solution, where the potential $V_{n}(x)$ with a selected
number $n$ does not coincide with any other potential $V_{m}(x)$ with
another number $m \ne n$ in this hierarchy. I.~e. the solution for
the general form of the superpotential of this hierarchy is not
reduced to a simple change of a form:
\begin{equation}
  W_{n+m}(x) = (-1)^{m} W_{n}(x).
\label{eq.4.1.10}
\end{equation}

\subsection{Potentials of Gamov's type 
\label{sec.4.2}}

\begin{figure}[ht]
\centerline{
\includegraphics[width=57mm]{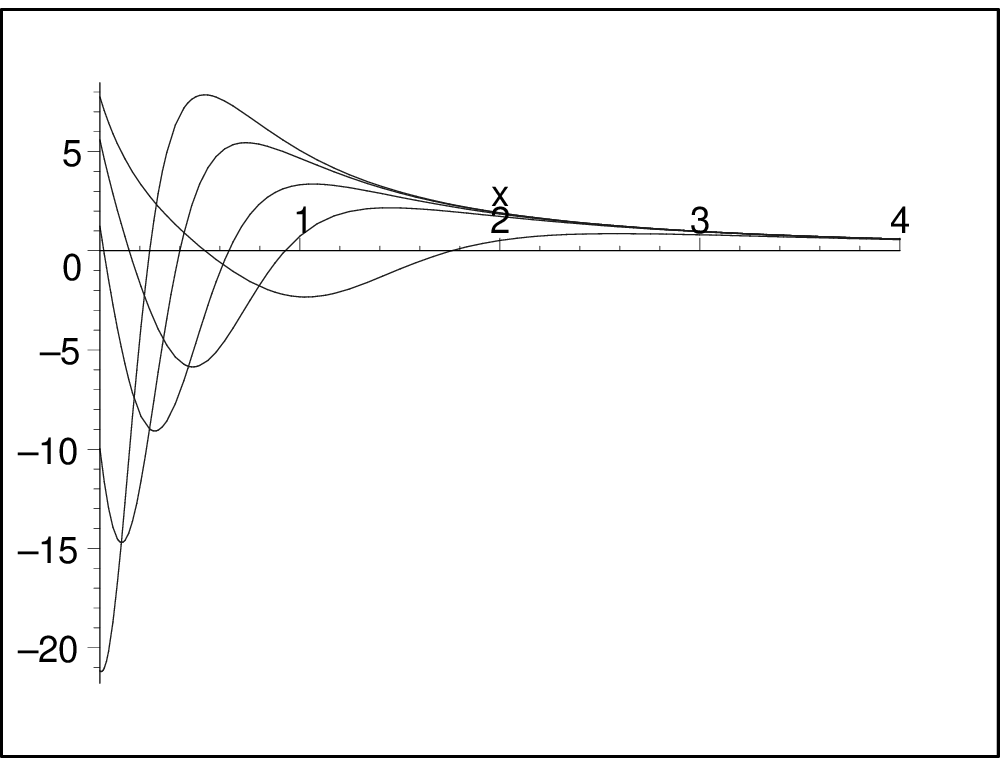}
\includegraphics[width=57mm]{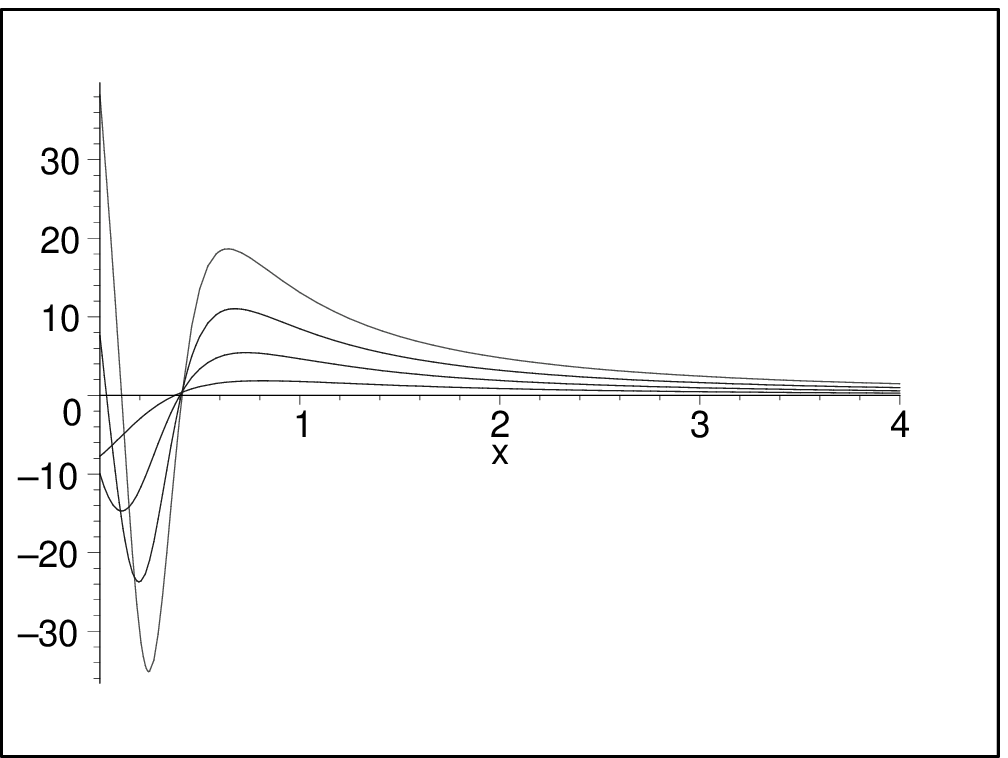}}
\caption{
A dependence of the potential $V_{n+1}(x)$ on
$C_{1}$ and $\gamma$:
(a) --- the barrier maximum and the hole minimum of this potential
are changed along the axis $x$ at the change of $C_{1}$ 
(at $C_{1} = 0.01, 0.1, 0.3, 1.0, 2.5$, $\gamma=6$, $x_{0}=0.5$);
(b) --- the barrier maximum of this potential practically does not
changed along the axis $x$ at the change of $\gamma$ 
(at $C_{1} = 1$, $\gamma=2, 6, 12, 20 $, $x_{0}=0.5$).
\label{fig.51}}
\end{figure}

\begin{figure}[ht]
\centerline{
\includegraphics[width=80mm]{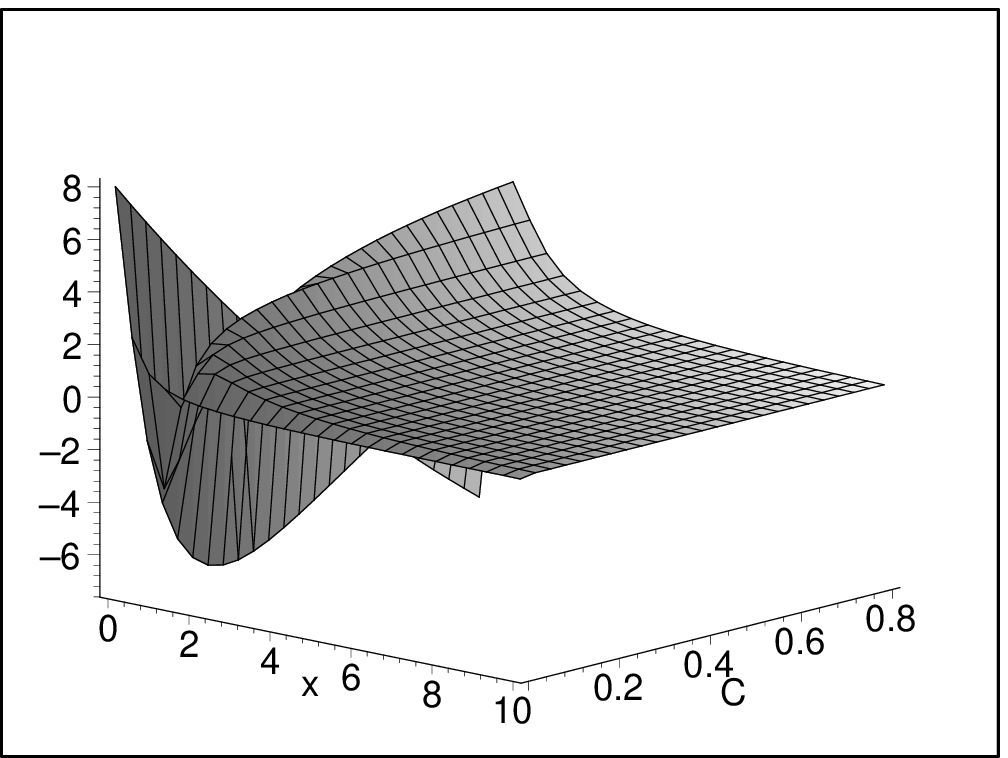}
\includegraphics[width=80mm]{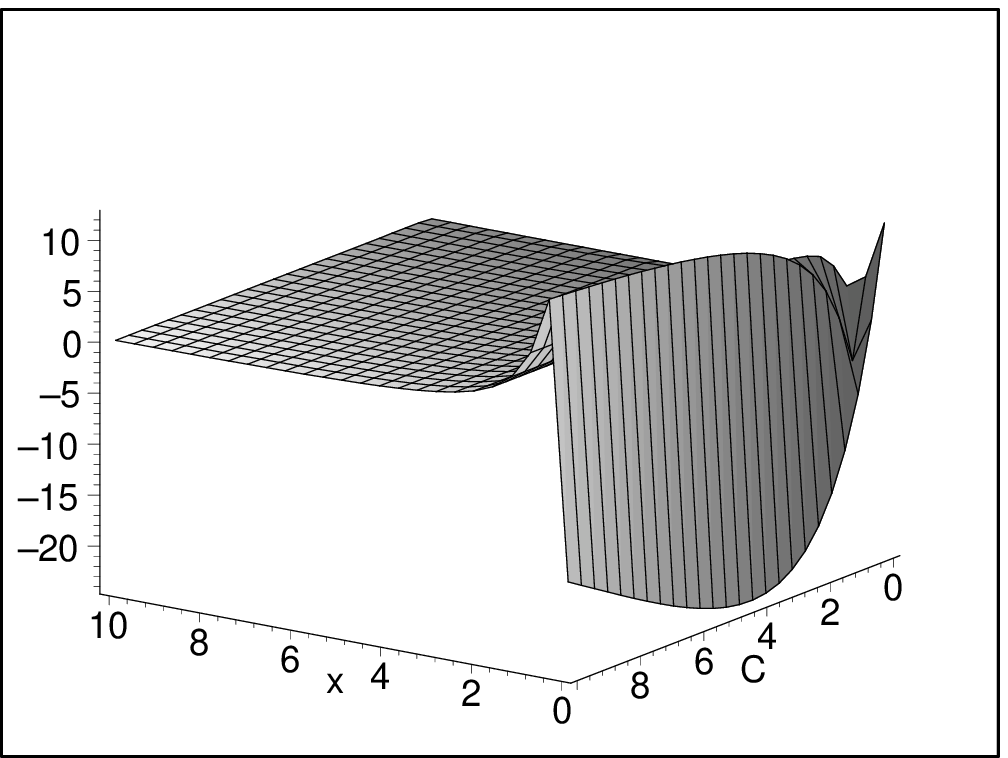}}
\caption{New exactly solvable potential $V_{n+1}(x)$
of Gamov's type (at $C_{n} \ne 0$).
Continuous change of its shape at variation of $C_{1}$
($\gamma=6$, $x_{0}=0.5$)
\label{fig.52}}
\end{figure}

Let's consider the superpotential (\ref{eq.4.1.4}) at
$C_{1} \ne 0$ and at $2\beta_{n} \ne \alpha$.
We shall find the potentials-partners, connected with it.
We shall study them in the positive semi-axis $x \ge 0$ only.
The potential
 $V_{n+1}(x)$ with the number $n+1$ is shown in
Fig.~\ref{fig.51}~(a, b) for selected values of the parameters
$C_{1}$ and $\gamma$.
%
%
%
%
From these figures one can see, that this potential has one hole and
one barrier, after which it falls down monotonously to zero with
increasing of a spatial coordinate. In its behavior this potential
looks qualitatively (at limit $x_{0} \to +0$) like potentials, 
used in theory of nuclear
collisions for description of a scattering of particles on spherically
symmetric nuclei, and also for description of decays and synthesis
of nuclei with spherical shape.
In Fig.~\ref{fig.52} a continuous deformation of the shape of this
potential at a change of the parameter $C_{1}$ is shown (here, one
can analyze the deformation of the barrier and the hole of this
potential).
%
%
We see that at $C_{1} \to 0$ this potential tends to the inverse
power potential. One can make sure in this also by a direct calculation
of the superpotential (\ref{eq.4.1.4}), which tends to the form
(\ref{eq.4.1.3}), and the potential tends to the form (\ref{eq.4.1.8}).
At $C_{1} \ne 0$ the new potential is not inverse power one and,
therefore, is the new exactly solvable solution. We note that one
can consider it as an isospectral expansion of the early found
potentials of the inverse power hierarchy (\ref{eq.4.1.8}).

\begin{figure}[ht]
\centerline{
\includegraphics[width=57mm]{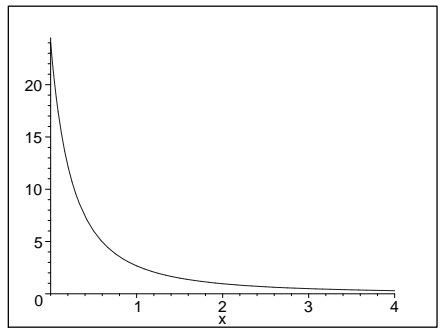}}
\caption{
Inverse-power potential $V_{n}(x)$
(at $C_{1}=1$, $\gamma=6$ and $x_{0}=0.5$).
It does not depend on change of parameters $C_{1}$ and $\gamma$.
\label{fig.53}}
\end{figure}

Another potential $V_{n}(x)$ SUSY-partner with a smaller number
$n$ is shown in Fig.~\ref{fig.53}. We see that it has the inverse
power dependence on the spatial coordinate. One can make sure that
it is not deformed with the change of the coefficient $C_{1}$ and
it belongs to the hierarchy of the inverse power potentials
(\ref{eq.4.1.8}), studied in the previous section.
%


If to use the value of the parameter $\gamma$ from the succession
(\ref{eq.4.1.9}), then the potential $V_{n}(x)$ becomes reflectionless,
in accordance with the \emph{Rule}, formulated above.

So, we have found the exactly solvable reflectionless potentials,
which in their shape look qualitatively 
(at $x_{0} \to 0$ and $x_{0} \ne 0$),
like the potentials of the scattering in description of the 
particles on nuclei of the spherical type, and also in 
description of the decays and synthesis of the nuclei 
of the spherical type.

If to study properties of the found potentials at the limit
$x_{0} \to 0$, then one can conclude that these potentials are
the potentials-partners to $\delta$-function, which has a peculiarity
at zero. Because of this the found potentials must to have the
peculiarity at zero also, and it is not clear, which penetration
they have.
However, a preliminary analysis has shown, that practically all part
of such found potential without an enough small neighborhood at zero
does not influence on the propagation of the particle in its field,
because of it must be reflectionless. Then at a spherically
symmetric consideration of the scattering of the particle in the
field of this potential, a radial component of which represents the
found potential, the particle propagates through it without the
smallest reflection, and, therefore, without change of the angle
of the propagation (except for an exact fall into a center). In
this case one can conclude, that the found potential (in spite of it
has the barrier) is reflectionless for the incident particle. If to
use it for description of the scattering of the particle on a
quantum system with the barrier, then one can conclude that such
system must be \emph{unvisible} for the incident particle, in spite
of it has the scattering barrtier.
Therefore, one can name the found potentials as \emph{the
reflectionless potentials of Gamov's type}.
At first time these potentials were opened in
\cite{Maydanyuk.2005.APNYA}.

\section{Conclusions
\label{sec.conclusions}}

In the finishing, we note main conclusions and new results.
\begin{itemize}
\item
We construct a hierarchy (defined at all coefficients $C_{n}=0$),
where the lowest energy level of a Hamiltonian with an arbitrary
number coincides with the lowest energy levels for Hamiltonians
with other numbers.
It has shown, that potentials of such hierarchies (defined at
$C_{n}=E_{0}^{(n)}=0$) with continuous energy spectra correspond to
real processes of scattering between particles and nuclei (because
of the scattering potentials in the asymptotic spatial regions tend
to zero) and, therefore, such hierarchies can be more useful from a
physical point of view in study of the collisions between the
particles and the nuclei in a comparison on the most widely used
type of the hierarchy, where $C_{n}=E_{0}^{(n)} \ne 0$ (for example,
see p.~287--289 in~\cite{Cooper.1995.PRPLC}).

An application of such hierarchies in study of the reflectionless
potentials has substantiated from the physical point of view.

\item
It has shown, that the hierarchy, defined at $C_{n}=0$, establishes
an interdependence (on the basis of Darboux transformations) between
wave functions and levels of the continuous energy spectra of the
Hamiltonians-partners.
For the first time it has shown, that an application of boundary
conditions makes the interdependence between the wave functions and
the energy levels of the Hamiltonians-partners with the continuous
energy spectra unique, and a selection of the boundary conditions
does not influence on the interdependence between amplitudes of
transmission and reflection and between coefficients of penetrability
and reflection.

\item
A Rule of determination of new reflectionless potentials on the
basis of known ones has corrected (at first, it was formulated in
\cite{Maydanyuk.2005.APNYA}, p.~447, sec.~3).
A proof of its fulfillment has proposed.

\item
For the first time it has proved, that at a known partial solution
for a superpotential $\tilde{W}_{n}(x)$ with a selected number $n$
in the SUSY-hierarchy, defined at $C_{n}=0$, there is a partial
solution for a superpotential $\tilde{W}_{n+m}(x)$ with an arbitrary
number $n+m$, and these superpotentials are connected by (\ref{eq.3.2.2}).

\item
For the first time a general solution for a superpotential
$W_{n+m}(x)$ with an arbitrary number $n+m$ (and a general form of
potentials-partners, connected with this superpotential) in a
hierarchy, defined at $C_{n}=0$, has found at a supposition, that
only one partial solution for the superpotential $\tilde{W}_{n}(x)$
with the selected number $n$ in this hierarchy is known.

\item
For the first time a general solution for a superpotential with an
arbitrary number and a general form of potentials, connected with
it, in the hierarchy, defined at $C_{n}=0$ and where the potentials
have an inverse power dependence on a spatial coordinate (and where
tunneling is possible), i.~e. of the form
$V(x) = \pm \alpha / |x-x_{0}|^{n}$
(where $\alpha$ and $x_{0}$ are constants, $n$ is a natural number),
have obtained.
A general solution for the hierarchy of the reflectionless inverse
power potentials has found.
These solutions are determined uniquely on the basis of the only one
known solution for the superpotential with the selected number.
The found hierarchies can be interesting as an example of solution
of a known problem of search of the general form of the hierarchies,
defined at different values $C_{n}$ (see p.~374, sec.~13.1
in~\cite{Cooper.1995.PRPLC}).

\item
A consequent statement and an analysis (with substantiation of some
points, which are not taken into account in~\cite{Maydanyuk.2005.APNYA})
of the reflectionless potentials, which in the spatial semi-axis have
one hole and one barrier, after which they fall down to zero
monotonously, are presented (at first, these potentials are
considered in~\cite{Maydanyuk.2005.APNYA}).
Such potentials at their shape look qualitatively like scattering
potentials in two-particle description of collisions between
particles and nuclei or decay potentials in two-particle description
of decay of compound spherical nuclear systems.
One can name such potentials as \emph{the potentials of Gamov's type}.
\end{itemize}

We note that the found potentials of Gamov's type can be interesting
in that they are exactly solvable, at own shape they look
qualitatively like the scattering potentials at description of the
nuclear collisions, they have one barrier (and because of this,
tunneling in them is possible) and they can be reflectionless.

\section*{Acknowledgements}

Author expresses his deep gratitude to the organizers of the  
XXXII ITEP Winter School of Physics for warm hospitality.

\bibliographystyle{h-physrev4}
\bibliography{Rnucl_En}

\newpage
\listoffigures

\newpage

\end{sloppypar}
\end{document}